\documentclass[twocolumn,amsmath,amssymb,prl,showpacs]{revtex4}

\pdfoutput=1
\usepackage[english]{babel}
\usepackage[latin1]{inputenc}
\usepackage[T1]{fontenc}
\usepackage{amsmath}
\usepackage{verbatim}
\usepackage[pdftex]{graphicx}

\begin{document}

\bibliographystyle{aip}

\title{Optical-phonon resonances with saddle-point excitons in twisted-bilayer graphene}

\author{Ado Jorio*$^{1,2}$, Mark Kasperczyk$^{1}$,  Nick Clark$^{3}$, Elke Neu$^{4}$, Patrick Maletinsky$^{4}$, Aravind Vijayaraghavan$^{3}$, Lukas Novotny$^{1}$}

\date{\today}

\affiliation{
$^{1}$ Photonics Laboratory, ETH Z\"urich, 8093 Z\"urich, Switzerland\\
$^{2}$ Departamento de F\'isica, Universidade Federal de Minas Gerais, Belo Horizonte, MG, 31270-901, Brazil\\
$^{3}$ School of Materials and National Graphene Institute, The University of Manchester, Manchester M13 9PL UK.\\
$^{4}$ Department of Physics, University of Basel, Klingelbergstrasse 82, CH-4056 Basel, Switzerland\\
$^{*}$ Correspondence should be addressed to \textbf{adojorio@fisica.ufmg.br} \\
}

\begin{abstract}
Twisted-bilayer graphene (tBLG) exhibits van Hove singularities in the density of states that can be tuned by changing the twisting angle $\theta$. A $\theta$-defined tBLG has been produced and characterized with optical reflectivity and resonance Raman scattering. The $\theta$-engineered optical response is shown to be consistent with persistent saddle-point excitons. Separate resonances with Stokes and anti-Stokes Raman scattering components can be achieved due to the sharpness of the two-dimensional saddle-point excitons, similar to what has been previously observed for one-dimensional carbon nanotubes. The excitation power dependence for the Stokes and anti-Stokes emissions indicate that the two processes are correlated and that they share the same phonon.

\,

\textbf{KEYWORDS}: graphene, Raman scattering, optical reflectivity

\end{abstract}


\maketitle

\begin{figure}[t]
\centering
\centerline{\includegraphics[width=9cm,clip]{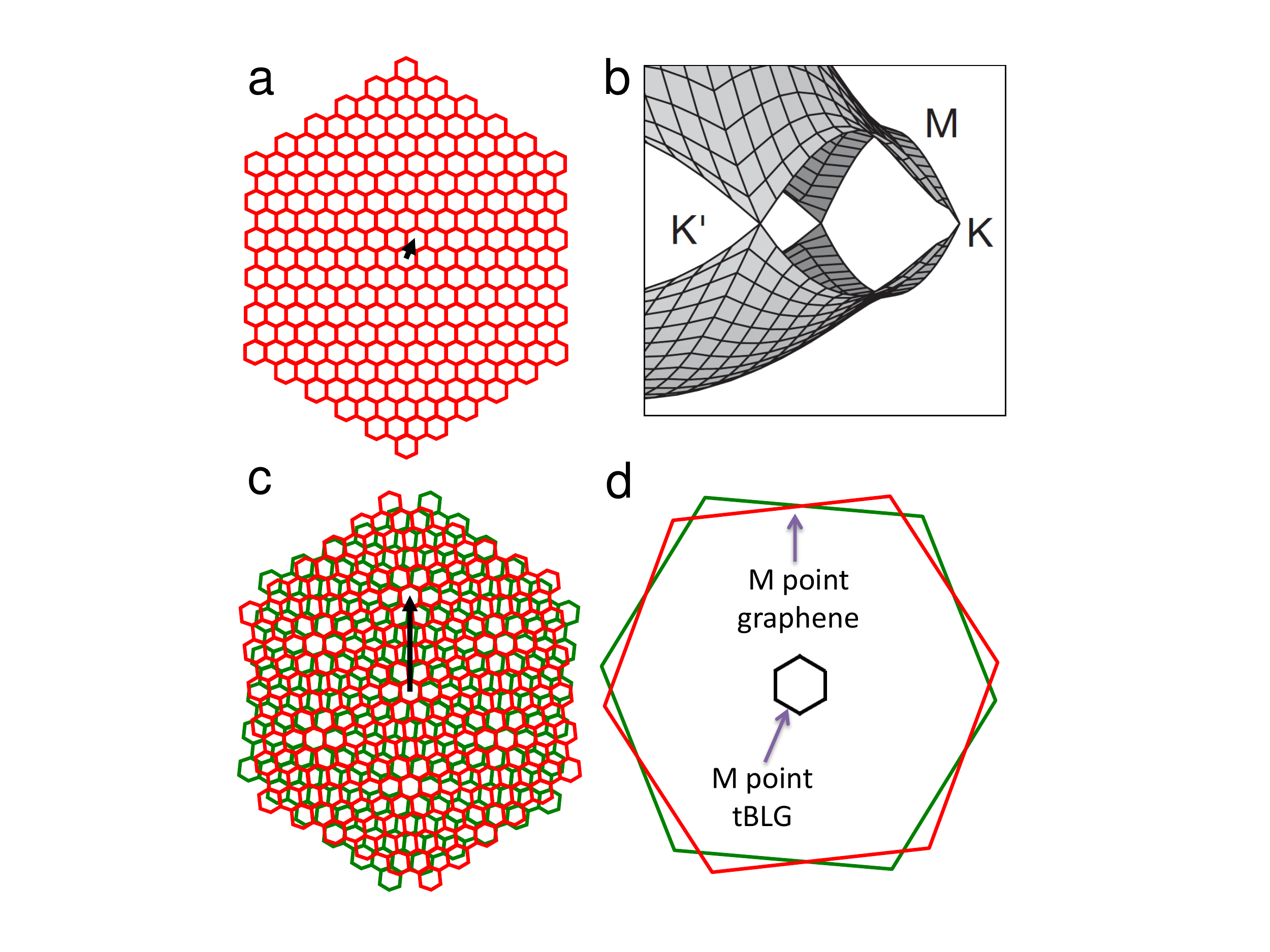}}
\caption{(a) Illustration of a single-layer graphene in real space, and (b) the dispersion of the $\pi$ and $\pi^*$ electrons at the edge of the Brillouin zone, close to the Dirac (K and K$^{\prime}$) and saddle (M) points \cite{raman-book}. (c) Illustration of a tBLG showing the superstructure Moir\'e pattern that defines the increased lattice periodicity (arrow). (d) The first Brillouin zones of the two single-layer graphenes (green and red) and of the tBLG (black). The location of the M points are indicated.}
\label{figMpoint}
\end{figure}

Saddle points in the electron dispersion of two-dimensional systems generate logarithmic diverging van-Hove singularities (vHs) in the density of states (DOS) \cite{vHs}. In single-layer graphene (Fig.\,\ref{figMpoint}a), the vHs occur at the M point for both the valence ($\pi$) and conduction ($\pi^*$) bands (Fig.\,\ref{figMpoint}b), and consequently in the joint density of sates (JDOS) for optical properties. The JDOS vHs falls into the ultra-violet frequency region in graphene \cite{raman-book}, as well as the related peak in the optical conductivity ($E_{\rm opt}$), which is redshifter from the JDOS vHs due to excitonic effects \cite{heinz-Mpoint}.

Twisted-bilayer graphene (tBLG) are formed by adding a twist angle $\theta$ between two graphene layers on top of each other~\cite{mele2010,morell10,li2010,jorio2013}. The expansion of the real lattice parameter when building a superstructured tBLG (see arrow in Fig.\,\ref{figMpoint}c) generates a shrinking of the reciprocal space (Fig.\,\ref{figMpoint}d), thus bringing the Dirac (K and K$^{\prime}$) points close together and moving them closer to the M saddle point. Consequently, the vHs will shift in energy, thereby redshifting the JDOS vHs to visible frequencies and even into the far-infrared, depending on the twisting angle $\theta$. This effect has been measured in tBLG using scanning tunneling spectroscopy \cite{li2010} and it has been explored using Raman spectroscopy \cite{jorio2013,ni2009,gupta2010,carozo2011,righi2011,havaner2012,kim2012,sato2012,campos-delgado2013,carozo2013}.

Up to date, the studied tBLGs have been produced randomly, i.e. without controlling the twist angle $\theta$. Furthermore, the generated 2D vHs have been directly measured only by scanning tunneling spectroscopy\,\cite{li2010}, which is limited to energies near the Fermi level and is usually more invasive and less accurate than optical measurements. Here we report the production and optical characterization of a tBLG with a previously selected $\theta$ angle, chosen specifically to explore Stokes and anti-Stokes resonance Raman scattering effects. By mode-engineering the system to work in resonance with the anti-Stokes line, we observe a quadratic power dependence of the anti-Stokes process. This power dependence is indicative of Stokes-induced anti-Stokes scattering in which the phonon generated by the Stokes scattering is consumed by the anti-Stokes process.

\begin{figure}[t]
\centering
\centerline{\includegraphics[width=9cm,clip]{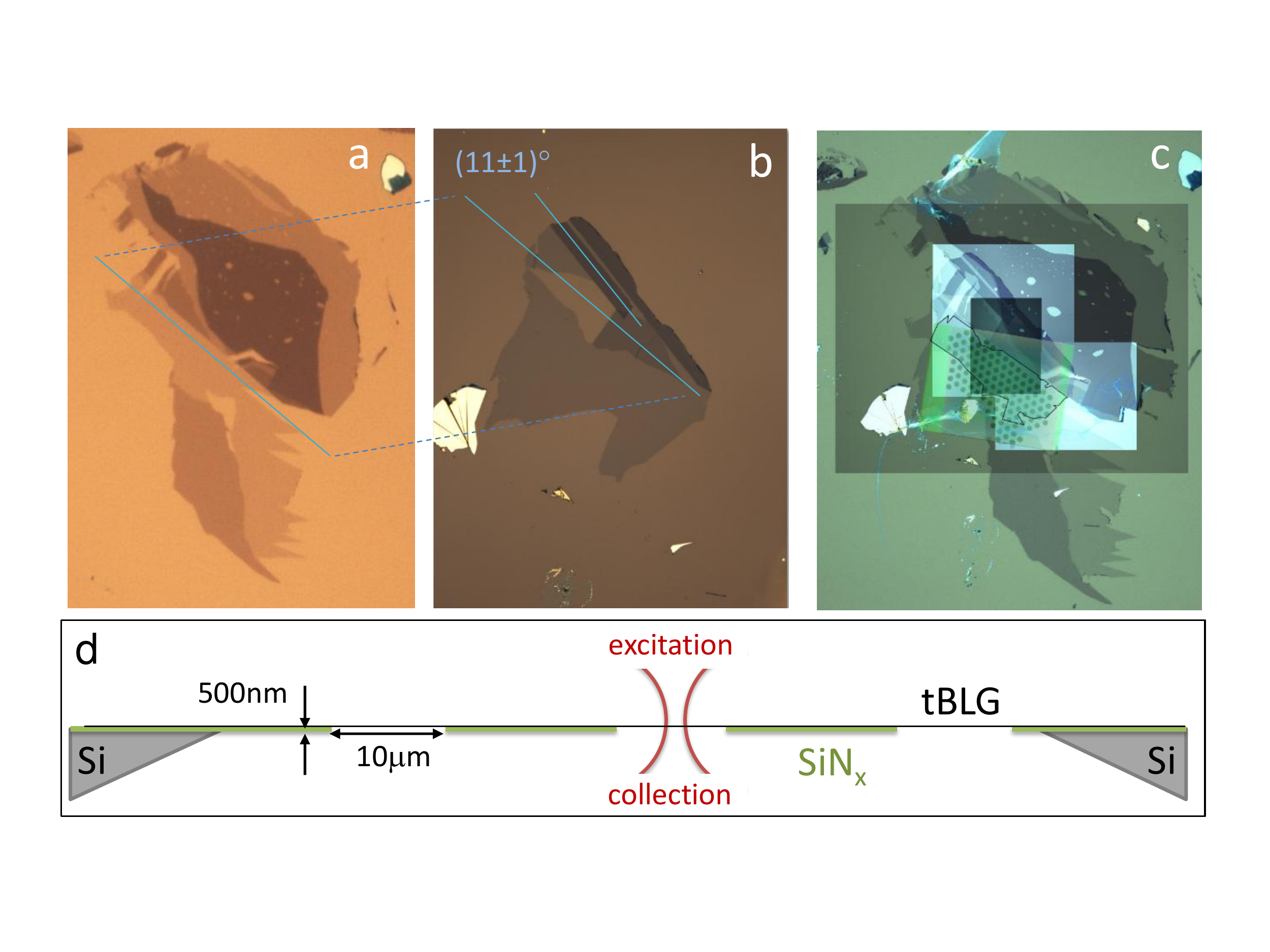}}
\caption{(a) and (b) show the optical images of two exfoliated graphene samples with well-defined edges (see light blue traces). The two graphene samples were superimposed with a well-defined angle $\theta = (11 \pm 1)^{\circ}$. The twisted bilayer graphene (tBLG) is then deposited on top of a Si substrate with a 500\,nm thick layer of SiN$_x$, perforated with $10\,\mu m$-diameter holes (optical image in (c) and schematics in (d)).}
\label{figsample}
\end{figure}

Resonance Raman studies show that the peak in the optical conductivity is given by $E_{\rm opt} = E_0\sin{(3\theta)}$, with $E_0 = 3.9{\rm eV}$ obtained experimentally \cite{carozo2013}. By choosing $\theta  = 11.3^{\circ}$, we expect $E_{\rm opt} = 2.175{\rm eV}$. Our tBLG sample was prepared from mechanically-exfoliated single-layer graphene flakes with well-defined edges (see Fig.\,\ref{figsample}a,b). After a first graphene flake had been transferred on top of a solid SiO$_2$ surface, a second flake is put down, rotated with respect to the first by an angle $\theta = 11 \pm 1^{\circ}$. Then the $\theta = 11 \pm 1^{\circ}$ tBLG was transferred onto a 500\,nm thick SiN$_x$ membrane perforated with holes of 10\,$\mu$m diameter (see optical image in Fig.\,\ref{figsample}c and schematics in Fig.\,\ref{figsample}d). Notice that while well-defined edges are a sufficient indication of armchair or zigzag atomic structures, they cannot be differentiated optically, and the procedure here has 50\% chance of generating a ($60^{\circ}-\theta$) angle, rather than the wanted $\theta$. We also fabricated a reference sample with $\theta = 0^{\circ}$ by directly depositing a mechanically exfoliated AB-stacked bilayer graphene (AB-BLG). All experiments were performed in the freely suspended region of the graphene tBLG, that is, in the holes of the SiN$_x$ membrane. This excludes any substrate-related background scattering and artifacts.

\begin{figure}[t]
\centering
\centerline{\includegraphics[width=7cm,clip]{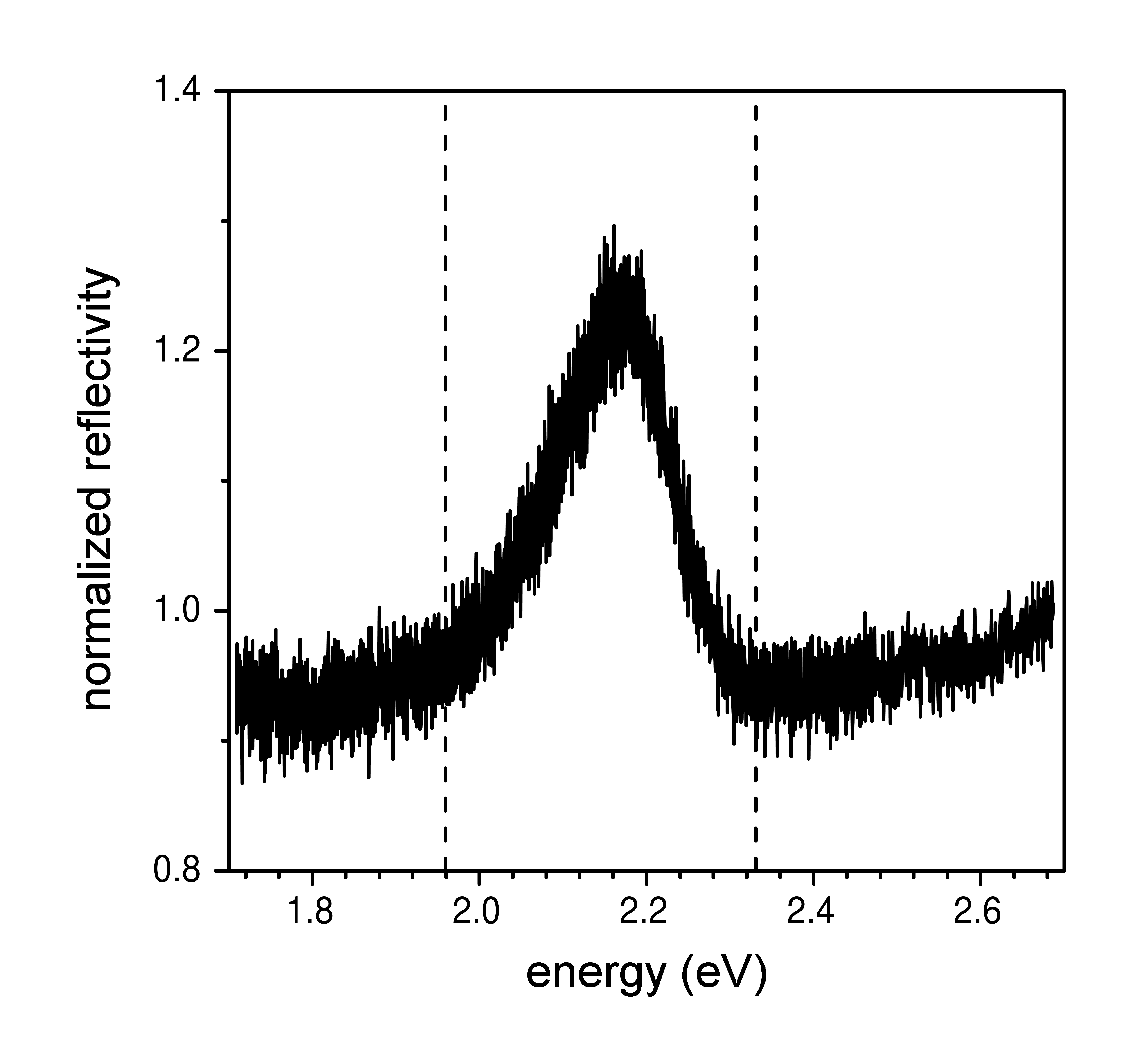}}
\caption{Reflectivity of the tBLG normalized by the reflectivity of AB-BLG. The vertical dashed lines show the locations of the excitation lasers $E_{\rm laser} =  1.96$\,eV and $E_{\rm laser} =  2.33$\,eV.}
\label{figreflectivity}
\end{figure}

Figure\,\ref{figreflectivity} shows the reflectivity of the tBLG shown in Fig.\,\ref{figsample}c, measured with a supercontinuum laser and normalized to the response of the AB-BLG. The tBLG reflectivity peak is observed at $E_{\rm opt} = 2.18{\rm eV}$, in excellent agreement with the prediction for $\theta = 11.3^{\circ}$ \cite{carozo2013}. The asymmetric reflectivity lineshape observed in the ultra-violet for graphene has been identified as a result of the convolution between a well-defined (excitonic) transition with a continuum density of electronic states, which results in a Fano resonance feature \cite{heinz-Mpoint}. The asymmetric lineshape shown in Fig.\,\ref{figreflectivity} is fully consistent with this picture, showing that many-body effects leading to the formation of two-dimensional saddle-point excitons are persistent, remaining in the modified structure of tBLG.

\begin{figure}[t]
\centering
\includegraphics[width=7cm,clip]{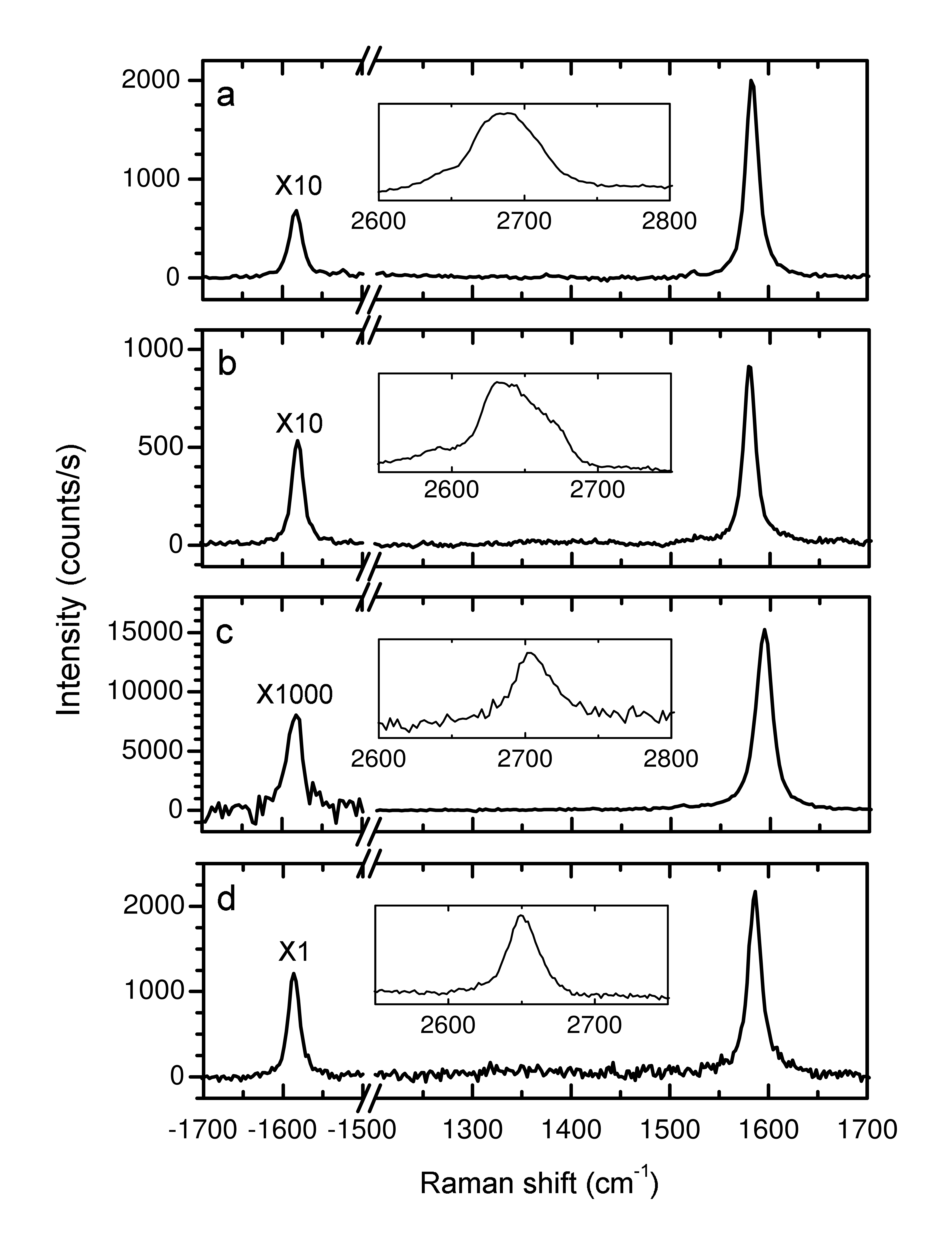}
\caption{Stokes and anti-Stokes Raman spectra of AB-BLG (a,b) and tBLG (c,d). (a) and (c) have been obtained with excitation at $E_{\rm laser} =  1.96$\,eV, while (b,d) have been obtained with excitation at $E_{\rm laser} =  2.33$\,eV. All spectra have been acquired with $P_{\rm laser}\sim$\,5\,mW reaching the sample through an air objective (NA = 0.8). The signal is collected in transmission by another air objective with larger NA (= 0.9). The Stokes spectra are displayed down to 1200\,cm$^{-1}$ to highlight the absence of the disorder-induced (D) band. The insets show the respective second-order G$^{\prime}$ (or 2D) bands, characteristic for AB-stacked (a,b) and misoriented (c,d) bilayer graphene. The anti-Stokes signals are magnified for clarity (magnification indicated on top of the respective peaks). All spectra have been corrected to account for detector (charged coupled device from Excelon - Pixis 100B) and spectrometer (Acton SP2300 from Princeton Instruments, equipped with a 600\,groves/mm grating blazed at 750\,nm) nominal efficiencies.}
\label{figRaman}
\end{figure}

Having properly characterized the energy of the saddle-point exciton ($E_{\rm opt} = 2.18{\rm eV}$), we can now explore different phenomena in the resonance Raman scattering.
Figure\,\ref{figRaman} shows the anti-Stokes and Stokes G band Raman spectra for AB-BLG (a,b) and tBLG (c,d). In Fig.\,\ref{figRaman}a,c the samples were excited with the HeNe laser at $E_{\rm laser} =  1.96$\,eV ($\lambda_{\rm laser} = 632.8$\,nm), while in Fig.\,\ref{figRaman}b,d the samples were excited with a frequency doubled YAG laser at $E_{\rm laser} =  2.33$\,eV ($\lambda_{\rm laser} = 532.0$\,nm). These two laser energies are marked in Fig.\,\ref{figreflectivity} by the vertical dashed lines. Notice that the two lasers are off resonance with respect to the saddle-point exciton, $E_{\rm laser}$ being blue- or red-shifted from $E_{\rm opt}$ by approximately the G band phonon energy ($E_{\rm G} = 0.2$\,eV).

In the AB-BLG case (Fig.\,\ref{figRaman}a,b), the Stokes and anti-Stokes signals do not change significantly when changing the excitation laser energy. The spectra obtained with $E_{\rm laser} =  1.96$\,eV and $E_{\rm laser} =  2.33$\,eV are similar (see Fig.\,\ref{figRaman}a,b), and the $I_{aS}/I_S$ ratio is basically unchanged. However, in the case of tBLG, the spectra change significantly when changing the excitation laser energy. Figure\,\ref{figRaman}c shows a Stokes signal $\sim$10 times stronger than the average signal obtained from the AB-BLG. Figure\,\ref{figRaman}d, on the other hand, shows an anti-Stokes signal $\sim$10 times stronger than in the AB-BLG. The $I_{aS}/I_S$ ratio thus changes by two orders of magnitude when changing the measurements with $E_{\rm laser} =  1.96$\,eV and $E_{\rm laser} =  2.33$\,eV. This result happens because, while the $E_{\rm laser} =  1.96$\,eV laser leads to a resonance with the anti-Stokes photon emission, the $E_{\rm laser} =  2.33$\,eV laser rise to a resonance with the Stokes photon emission. Similar results were observed in one-dimensional (1D) carbon nanotubes \cite{agsf2001}, where sharp resonances are characteristics of the 1D band structure \cite{jorio2001}. Interestingly, similar effect can be observed in a two-dimensional graphene system due to the sharp saddle-point exciton resonance.

\begin{figure}[t]
\centering
\includegraphics[width=7cm,clip]{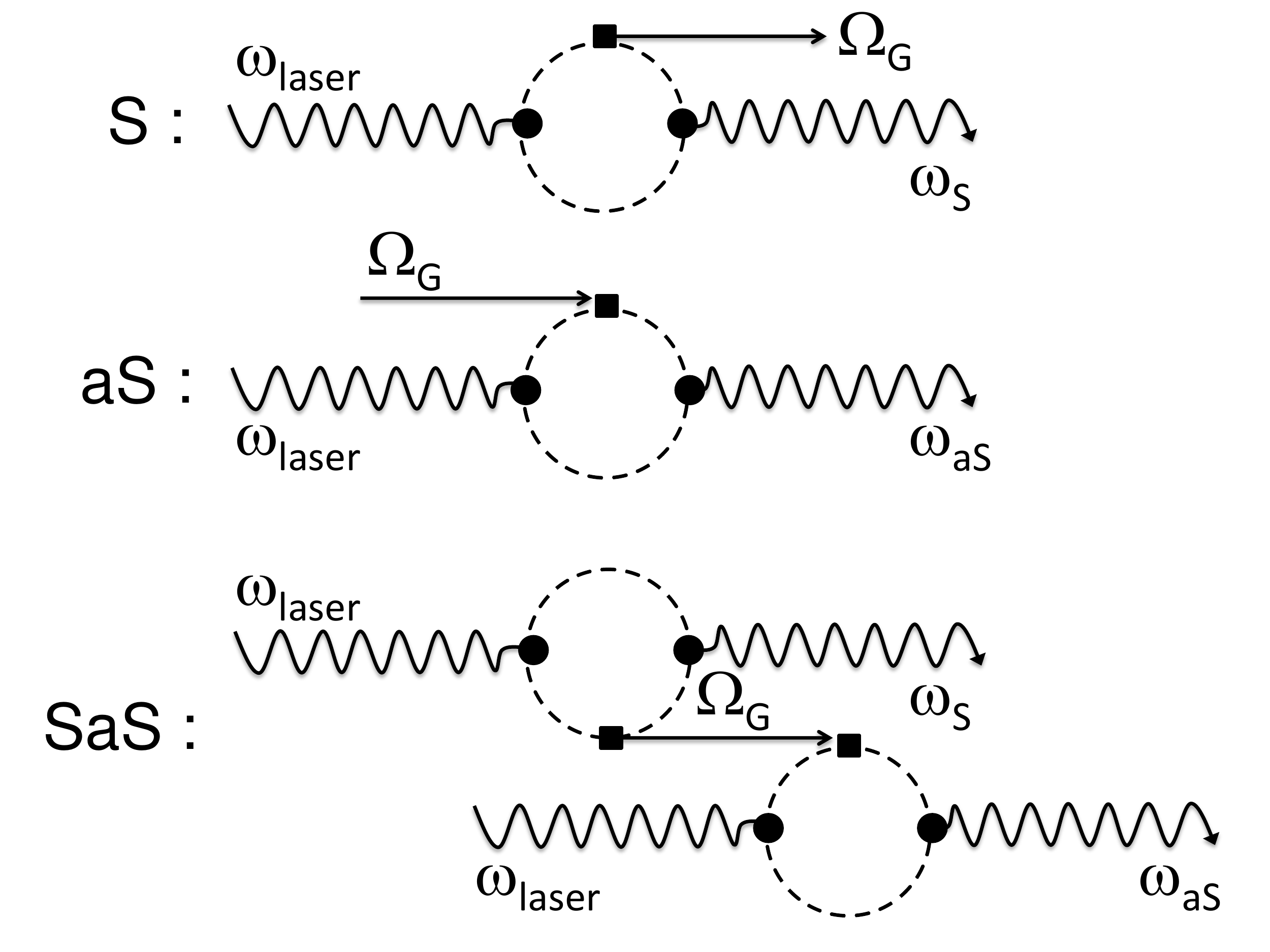}
\caption{Feynman diagrams for three different Raman scattering processes. Wavy and straight arrows stand for photons and phonons, respectively, while the dashed circles
represent electron-hole pairs. Black dots and black squares represent electron-photon and electron-phonon interactions. S stands for Stokes Raman process, aS for anti-Stokes
Raman process, and the last diagram, named here SaS, represents the generation of a Stokes and anti-Stokes photon pair via creation and annihilation of the same phonon.}
\label{figFeinman}
\end{figure}

Notice that in Figure\,\ref{figRaman}d, the anti-Stokes G band signal intensity is close to the Stokes intensity. Although similar result has been observed in carbon nanotubes before \cite{agsf2001,jorio2001}, an important question remains: where do these high energy G band phonons, which are being annihilated in the highly efficient resonant anti-Stokes emission, come from? To answer this question, we refer to the Feynman diagrams shown in Figure\,\ref{figFeinman}. The S diagram shows the Stokes Raman scattering event where a phonon and a Stokes-shifted Raman photon are generated. The aS diagram shows the anti-Stokes Raman scattering process, where one existing phonon is annihilated, generating the anti-Stokes shifted Raman photon. The aS process depends on the previous existence of the phonon in the lattice. In the case of graphene, $E_{\rm G} = 0.2$\,eV is much higher then the thermal (room temperature) energy. The last diagram in Figure\,\ref{figFeinman}, which we call here the Stokes-anti-Stokes (SaS) event, describes a non-linear process where a phonon created by the Stokes process is subsequently annihilated in the anti-Stokes process. The Stokes and anti-Stokes photons emerging from the SaS event can be correlated and, as discussed by D. N. Klyshko \cite{klyshko}, the character of this correlation can be continuously varied from purely quantum to purely classical. In order to observe the SaS scattering, we need to efficiently consume photo-generated phonons in the anti-Stokes process. Otherwise, the phonons will decohere via scattering. We also need high energy optical phonons that are unlikely to be populated at room temperature. These constrains are achieved in our $\theta$-defined tBLG system.

To differentiate the anti-Stokes signal originated from the SaS event from that produced by the usual aS process, we analyze the excitation laser power ($P_{\rm laser}$) dependence of the anti-Stokes Raman intensity ($I_{aS}$). Considering both the aS and SaS events in Fig.\,\ref{figFeinman} separately, $I_{aS}$ assumes the form:
\begin{equation}
  \label{eq1}
I_{aS} = P_{\rm laser}\{C_{aS} \eta + C_{SaS}P_{\rm laser}\} \, ,
\end{equation}
where $C_{aS}$ and $C_{SaS}$ are constants and $\eta = 1/[\exp{(\hbar\omega_{phonon}/k_BT)}-1]$ is the temperature dependent Bose-Einstein phonon distribution function. In the thermal regime, the dependence of $I_{aS}$ on $P_{\rm laser}$ reflects the material temperature through $\eta$.
In the regime where the SaS event dominates, $I_{aS}$ is proportional to $P_{\rm laser}^2$.

\begin{figure}[t]
\centering
\includegraphics[width=7cm,clip]{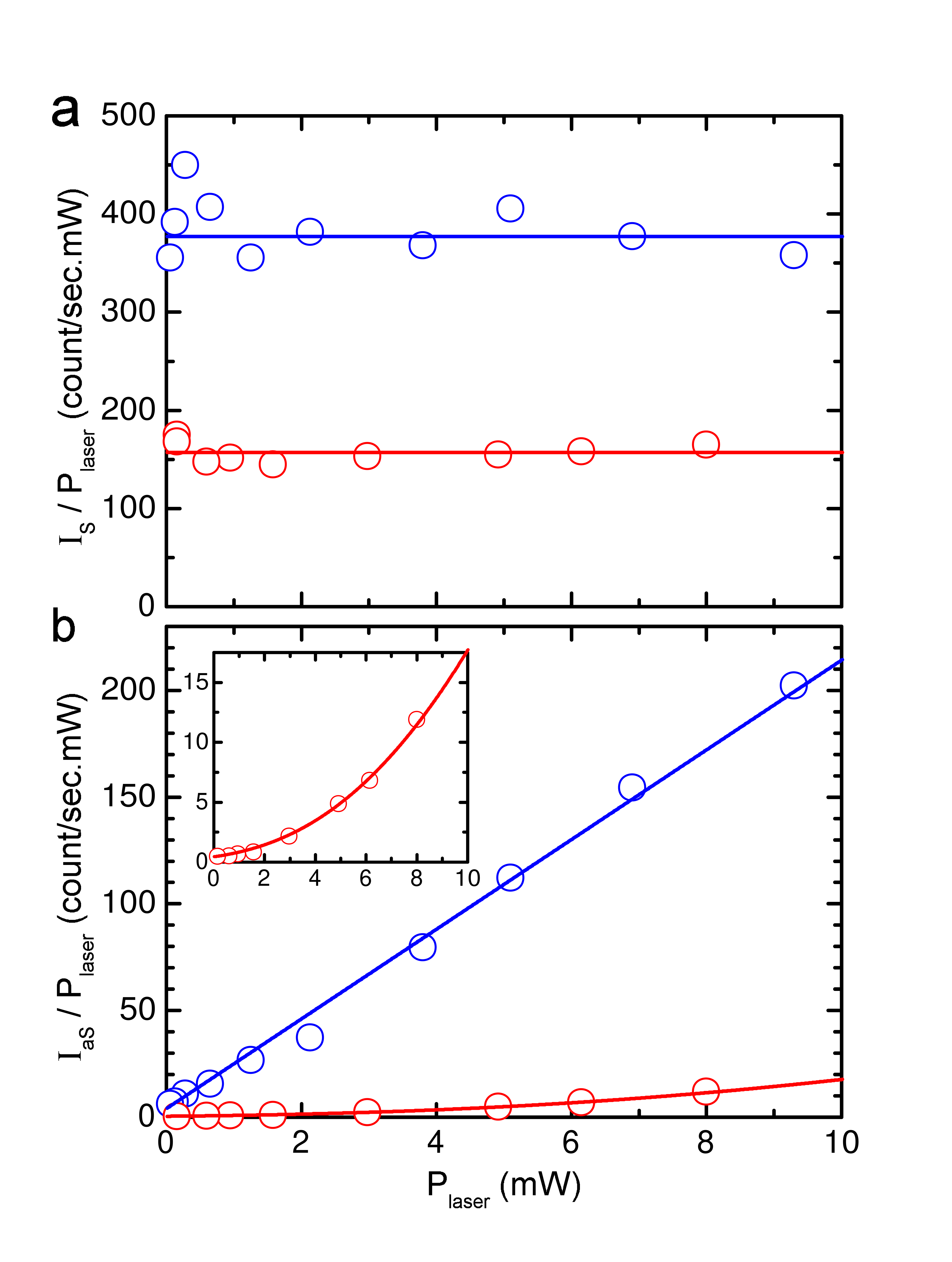}
\caption{Power dependence of (a) Stokes ($I_S$) and (b) anti-Stokes ($I_{aS}$) Raman intensities for the tBLG (blue) and AB-BLG (red) obtained with $E_{\rm laser} =  1.96$\,eV. The G band peaks were fit with a Lorentzian function. The Raman intensities were obtained from the peak integrated areas and then normalized by the laser power ($I_{S}/P_{laser}$ and $I_{aS}/P_{laser}$) to clearly demonstrate the difference in behavior. Circles are experimental data and lines are fitting functions according to Eq.\,\ref{eq1} (see text for fitting values). The inset in (b) is a magnified version of the AB-BLG data.}
\label{figanal}
\end{figure}

Figure\,\ref{figanal} shows the power dependence of the Stokes (a) and anti-Stokes (b) signals from the tBLG (blue) and from the AB-BLG (red), when excited with $E_{\rm laser} =  1.96$\,eV. The intensities are normalized by the excitation laser power. Figure\,\ref{figanal}a shows that the Stokes signal scales linearly with excitation power for both samples, which is indicative for the spontaneous scattering regime. However, the anti-Stokes Raman signals of the two samples show markedly different behaviors (Fig.\,\ref{figanal}b).

The anti-Stokes intensity of the tBLG excited at $E_{\rm laser} =  1.96$\,eV shows a quadratic power dependence within the measured range (0-10\,mW, blue data in Fig.\,\ref{figanal}b). The data can be accurately fitted by Eq.\,\ref{eq1} considering $\eta$ constant at $T = 295$\,K, with $C_{aS} = 9.1$\,W$^{-1}$s$^{-1}$ and $C_{SaS} = 2.3\times 10^{-9}$\,W$^{-2}$s$^{-1}$. For comparison, the power-normalized anti-Stokes intensity from the AB-BLG reference standard shows a much less pronounced $I_{aS}/P_{\rm laser}$ enhancement within the measured range (see red data in Fig.\,\ref{figanal}b), with a clear non-linear power dependence behavior (see inset). The AB-BLG $I_{aS}/P_{\rm laser}$ data can be fit with $C_{aS} = 1.0$\,W$^{-1}$s$^{-1}$, assuming that the temperature depends linearly on the excitation laser power as $T = 295+C_TP_{laser}$, with $C_T = 27$\,mW$^{-1}$K and $C_{SaS} = 0$. According with this latest equation, for $P_{\rm laser} > 10$,mW the AB-BLG temperature crosses 600$^{\circ}$C. The AB-BLG sample burns for laser powers above 12\,mW, while the tBLG stands much higher powers, evidencing the cooling process induced by the efficient anti-Stokes resonant emission. It is important to comment, however, that the presence of some heating in the tBLG or the SaS event in the AB-BLG reference sample cannot be ruled out from the results shown here, since a relatively small non-zero temperature increase for tBLG or $C_{SaS}$ term for AB-BLG can be added to the fitting curves without significantly compromising the fitting qualities.

The SaS process has been measured in diamond \cite{Lee2012}. The experiments were carried out in the femtosecond pump regime, where the extremely high number of photons per pulse significantly enhances the SaS probability. It is surprising that our results indicate the observation of the SaS event in graphene with a significantly low power from a CW laser. Lee {\it et al.} \cite{Lee2012} demonstrated the quantum nature of the SaS correlation in bulk diamond at room temperature by measuring a non-classical SaS second-order time correlation function $g^{(2)}$. This type of measurement has to be developed in graphene, so that the correlation nature of the SaS event, as proposed by D. N. Klyshko \cite{klyshko} and observed for bulk diamond \cite{Lee2012}, could be explored.

We expect the SaS event also to play a role in other low dimensional materials, like in carbon nanotubes. Steiner {\it et al.} \cite{steiner2007} observed laser-induced phonon cooling in single-wall carbon nanotubes by introducing them into an external optical cavity. They discussed how this scheme can be used to remove high energy hot phonons, which play an important role in the thermal and electrical properties of low-dimensional structures \cite{steiner2009,barreiro2009}. Here we designed a system that can perform like a cooling system, but the nanostructure itself is the cavity. Because of the M-point engineering scheme used to build the tBLG, the resonance can be tuned to work at any selected wavelength from the far-infrared up to the ultra-violet, in the amplifying or cooling mode, just by changing the tBLG angle $\theta$. Also, multiple twisted bilayer graphenes can be stacked (separated for example by hexagonal-BN) to enhance the efficiency further.

Finally, since Raman spectroscopy is established as an important tool to study and characterize nanostructures, our findings indicate that the SaS event has to be considered when using this technique to extract structural and transport properties of low-dimensional structures. In future work it will be interesting to study changes of the phonon (G band) linewidth and frequency as a function of detuning ($\omega_{\rm opt}-\omega_{\rm laser}$), which may change the influence of the Kohn anomaly \cite{piscanec} effects in the G band properties.

\section{Acknowledgement}

The authors strongly acknowledge valuable discussions with Dr. P. Bharadwaj and L. G. Can\c cado. A.J. acknowledges Inmetro for the background developed in tBLG, CAPES and ETH for financing his visit to ETH Zurich. L.N. thanks financial support by the Swiss National Science Foundation (SNF) through grant 200021-14 6358. A.V. thanks the Engineering and Physical Sciences Research Council (EPSRC) for financial support through grants EP/G035954/1 and EP/K009451/1.

\end{document}